# Modellvalidierung mit Hilfe von Quantil-Quantil-Plots unter Solvency II

**Dietmar Pfeifer**

**Version: 5.1.2020**

**Zusammenfassung** Nach etlichen Jahren Vorarbeit ist das Projekt Solvency II Anfang 2016 in den Ländern der Europäischen Union legislativ umgesetzt worden. Damit verbunden sind einige wesentliche Änderungen der jeweiligen nationalen Versicherungsaufsichtsgesetze. Ein neuer Aspekt hierbei ist die Vorschrift, potenzielle Abweichungen des Risikoprofils des Unternehmens von den Annahmen, die der Standardformel zur Berechnung des Solvency Capital Requirements (SCR) zugrunde liegen, zu analysieren und zu beurteilen. Für das Prämien- und Reserve-Risiko bzw. die zugehörigen Schaden-Kosten-Quoten wird dabei stillschweigend eine Lognormal­verteilung unterstellt. In dieser Arbeit wird ein einfaches, aber dennoch mathematisch korrektes Verfahren auf der Basis von Quantil-Quantil-Plots vorgestellt, mit dem eine solche Analyse durchgeführt werden kann.

**Model validation on the basis of quantile-quantile-plots under Solvency II**

**Abstract** After several years of development, the Solvency II-project has finally been set to work in the European Union with the beginning of the year 2016. This has caused massive changes in the regional legislative supervisory acts. One new aspect of regulation is the requirement of an analysis and judgement concerning possible deviations of the company's risk profile from the assumptions underlying the standard formula in Solvency II. In particular, for the reserve and premium risk and the corresponding combined ratios, resp. a lognormal distribution is implicitly assumed. In this paper, we present a simple, but nevertheless mathematically accurate method on the basis of quantile-quantile-plots which is suitable to perform such kind of analyses.

**0. Vorbemerkung**

In § 27 des neuen deutschen Versicherungsaufsichtsgesetztes heißt es: Zum Risikomanagement­system gehört eine unternehmenseigene Risiko- und Solvabilitätsbeurteilung, die Versicherungs­unternehmen regelmäßig sowie im Fall wesentlicher Änderungen in ihrem Risikoprofil unver­züglich vorzunehmen haben. … Die Risiko- und Solvabilitätsbeurteilung umfasst mindestens

1. eine eigenständige Bewertung des Solvabilitätsbedarfs unter Berücksichtigung des spezifischen Risikoprofils, der festgelegten Risikotoleranzlimite und der Geschäftsstrategie des Unternehmens,

D. Pfeifer (✉)
Institut für Mathematik, Schwerpunkt Versicherungs- und Finanzmathematik, Carl von Ossietzky Universität Oldenburg, Oldenburg, Deutschland
E-Mail: dietmar.pfeifer@uni-oldenburg.de

2. eine Beurteilung der jederzeitigen Erfüllbarkeit der aufsichtsrechtlichen Eigenmittelanforderungen, der Anforderungen an die versicherungstechnischen Rückstellungen in der Solvabilitätsübersicht und der Risikotragfähigkeit sowie
3. eine Beurteilung der Wesentlichkeit von Abweichungen des Risikoprofils des Unternehmens von den Annahmen, die der Berechnung der Solvabilitätskapitalanforderung mit der Standardformel oder mit dem internen Modell zugrunde liegen.

Konkret geht es hierbei um die Annahme einer Lognormalverteilung für das Prämien- und Reserve-Risiko bzw. für die jährlichen Schaden- bzw. Schaden-Kosten-Quoten (vgl. Dreher (2018), S. 912). Eine mathematisch korrekte Überprüfung dieses Sachverhalts kann eigentlich nur mit geeigneten statistischen Tests durchgeführt werden. In dieser Arbeit werden neue asymptotische Formeln hergeleitet, mit denen zum einen diese Annahme mit Hilfe von Quantil-Quantil-Plots überprüft werden kann, zum anderen erwartungstreue Schätzer der zugehörigen Lage-Skalen-Parameter für die logarithmierten Schaden- bzw. Schaden-Kosten-Quoten und damit Momente und Quantile (als Grundlage für das SCR, Solvency Capital Requirement) der entsprechenden Lognormalverteilung hergeleitet werden können. Quantil-Quantil-Plots haben darüber hinaus den Vorteil, die Testergebnisse auch graphisch veranschaulichen zu können, was insbesondere mathematisch weniger geschulten Mitarbeitern von Versicherungsunternehmen entgegenkommen dürfte.

**1. Einführung**

Graphische Methoden zur statistischen Analyse und Parameterschätzung in Lage-Skalen-Familien von Wahrscheinlichkeitsverteilungen haben eine lange Tradition, sie gehen auf sogenannte „Wahrscheinlichkeitspapiere" (Quantil-Quantil-Plots) zurück, die etwa ab dem Beginn des 20. Jahrhunderts vor allem im ingenieurwissenschaftlichen Kontext zur Anwendung kamen (vgl. den Übersichtsartikel von Cunnane (1978)). Lag der Schwerpunkt zunächst auf der Anpassung der Normalverteilung an hydrologische Beobachtungen, kamen später insbesondere Anpassungen an Extremwertverteilungen und andere Klassen von Wahrscheinlichkeitsverteilungen hinzu (Guo (1990)). Eine „optimale" Wahl der Plot-Positionen (auf der Abszisse) ist dabei eng verbunden mit der Berechnung von Erwartungswerten der geordneten Beobachtungen (sog. Ordnungsstatistiken, vgl. David und Nagaraja (2003)).

Später wurden die reinen Schätzverfahren für Lage- und Skalenparameter (oft identisch mit Erwartungswert und Streuung der zugrunde liegenden Verteilung) um geeignete Testverfahren erweitert, mit denen unabhängig von diesen Parametern das Vorliegen eines bestimmten Verteilungstyps überprüft werden kann. Ein interessanter Zugang besteht hier in der Verwendung (einer geeigneten Transformation) des empirischen Korrelationskoeffizienten aus dem Quantil-Quantil-Plot als Testgröße (vgl. Lockhart und Stephens (1998)).

## 2. Quantil-Quantil-Plots und Lage-Skalen-Familien: Schätzer

Betrachtet werden Risiken $X$ der Form $X = \mu + \sigma Z$ mit einem „Prototypen" $Z$ und stetiger, streng monotoner Verteilungsfunktion $F_Z$. Ziel ist die Schätzung der Parameter $\mu$ und $\sigma$ sowie die Überprüfung der Verteilungshypothese anhand von $n$ Beobachtungen. Bezeichnet dazu $X_{(k)}$ die $k$-te Ordnungsstatistik (d.h. den $k$-größten Wert) aus einer Reihe von unabhängigen Replikationen $X_1, \cdots, X_n$ von $X$, so trägt man im Quantil-Quantil-Plot die Größen $\left(Q_Z(u_k), X_{(k)}\right)$ mit der Quantilfunktion $Q_Z = F_Z^{-1}$ und geeigneten $u_k, k = 1, \cdots, n$ ab und ermittelt mittels linearer Regression die Ausgleichsgerade mit Achsenabschnitt $\hat{\mu}$ und Steigung $\hat{\sigma}$, die gegeben sind durch

$$\hat{\sigma} = \frac{\frac{1}{n}\sum_{k=1}^{n} X_{(k)} Q_Z(u_k) - \left(\frac{1}{n}\sum_{k=1}^{n} X_{(k)}\right) \cdot \left(\frac{1}{n}\sum_{k=1}^{n} Q_Z(u_k)\right)}{\frac{1}{n}\sum_{k=1}^{n} Q_Z^2(u_k) - \left(\frac{1}{n}\sum_{k=1}^{n} Q_Z(u_k)\right)^2} \quad \text{und} \quad \hat{\mu} = \frac{1}{n}\sum_{k=1}^{n} X_{(k)} - \frac{\hat{\sigma}}{n}\sum_{k=1}^{n} Q_Z(u_k) \quad (1)$$

(vgl. Fahrmeir et al. (2016), Abschnitt 3.6.2).

Die Bedeutung der Erwartungswerte der Ordnungsstatistiken für die Parameterschätzungen zeigt sich in der folgenden

Proposition: Mit der Wahl $u_k = F_Z\left(E\left(Z_{(k)}\right)\right)$ für $k = 1, \cdots, n$ sind $\hat{\sigma}$ und $\hat{\mu}$ erwartungstreue Schätzer für $\sigma$ und $\mu$.

Beweis: Aus der obigen Wahl folgt

$$Q_Z(u_k) = E\left(Z_{(k)}\right) = \frac{E\left(X_{(k)}\right) - \mu}{\sigma} \quad \text{bzw.} \quad E\left(X_{(k)}\right) = \mu + \sigma Q_Z(u_k) \text{ für } k = 1, \cdots, n, \quad (2)$$

wobei $Z_{(k)} = \frac{X_{(k)} - \mu}{\sigma}$ verteilungsmäßig der $k$-ten Ordnungsstatistik aus einer Reihe von unabhängigen Replikationen $Z_1, \cdots, Z_n$ von $Z$ entspricht. Es folgt

$$E(\hat{\sigma}) = \frac{\frac{1}{n}\sum_{k=1}^{n}E(X_{(k)})Q_Z(u_k) - \left(\frac{1}{n}\sum_{k=1}^{n}E(X_{(k)})\right)\cdot\left(\frac{1}{n}\sum_{k=1}^{n}Q_Z(u_k)\right)}{\frac{1}{n}\sum_{k=1}^{n}Q_Z^2(u_k) - \left(\frac{1}{n}\sum_{k=1}^{n}Q_Z(u_k)\right)^2}$$

$$= \frac{\frac{1}{n}\sum_{k=1}^{n}(\mu+\sigma Q_Z(u_k))Q_Z(u_k) - \left(\frac{1}{n}\sum_{k=1}^{n}(\mu+\sigma Q_Z(u_k))\right)\cdot\left(\frac{1}{n}\sum_{k=1}^{n}Q_Z(u_k)\right)}{\frac{1}{n}\sum_{k=1}^{n}Q_Z^2(u_k) - \left(\frac{1}{n}\sum_{k=1}^{n}Q_Z(u_k)\right)^2}$$

$$= \frac{\mu\cdot\frac{1}{n}\sum_{k=1}^{n}Q_Z(u_k) + \sigma\cdot\frac{1}{n}\sum_{k=1}^{n}Q_Z^2(u_k) - \mu\cdot\frac{1}{n}\sum_{k=1}^{n}Q_Z(u_k) - \sigma\cdot\left(\frac{1}{n}\sum_{k=1}^{n}Q_Z(u_k)\right)^2}{\frac{1}{n}\sum_{k=1}^{n}Q_Z^2(u_k) - \left(\frac{1}{n}\sum_{k=1}^{n}Q_Z(u_k)\right)^2} = \sigma \qquad (3)$$

und

$$E(\hat{\mu}) = \frac{1}{n}\sum_{k=1}^{n}E(X_{(k)}) - \frac{E(\hat{\sigma})}{n}\sum_{k=1}^{n}Q_Z(u_k) = \frac{1}{n}\sum_{k=1}^{n}(\mu+\sigma Q_Z(u_k)) - \frac{\sigma}{n}\sum_{k=1}^{n}Q_Z(u_k) = \mu, \qquad (4)$$

was zu zeigen war.

Bemerkung: Im Falle einer reinen Skalenfamilie (d.h. $\mu = 0$) gilt entsprechend

$$\hat{\sigma} = \frac{\frac{1}{n}\sum_{k=1}^{n}X_{(k)}Q_Z(u_k)}{\frac{1}{n}\sum_{k=1}^{n}Q_Z^2(u_k)} \qquad (5)$$

Auch dieser Schätzer ist erwartungstreu, wenn $u_k = F_Z(E(Z_{(k)}))$ für $k = 1,\cdots,n$ gewählt wird.

Beispiel: Standard-Exponentialverteilung: Hier ist $F_Z(x) = 1-e^{-x}$ für $x > 0$. Da die Ordnungsstatistiken unabhängige, exponentialverteilte Zuwächse besitzen, gilt

$$E(Z_{(k)}) = \sum_{i=1}^{k}\frac{1}{n+1-i} \text{ für } k = 1,\cdots,n, \qquad (6)$$

mit

$$u_k = F_Z(E(Z_{(k)})) = 1 - \exp\left(-\sum_{i=1}^{k}\frac{1}{n+1-i}\right), \ k = 1,\cdots,n. \qquad (7)$$

Approximativ ergibt sich hieraus mit $\sum_{j=1}^{m}\frac{1}{j} \approx \ln(m+1) + \gamma$ ($\gamma = 0{,}57721...$ ist die Euler-Konstante) die Beziehung

$$E\left(Z_{(k)}\right)=\sum_{i=1}^{k}\frac{1}{n+1-i}=\sum_{i=1}^{n}\frac{1}{i}-\sum_{i=1}^{n-k}\frac{1}{i}\approx \ln(n+1)-\ln(n+1-k)=\ln\left(\frac{n+1}{n+1-k}\right) \qquad (8)$$

bzw.

$$u_k = F_Z\left(E\left(Z_{(k)}\right)\right)=1-\exp\left(-\sum_{i=1}^{k}\frac{1}{n+1-i}\right)\approx \frac{k}{n+1},\ k=1,\cdots,n. \qquad (9)$$

Dies entspricht der „klassischen" Empfehlung von Weibull (1939), vgl. auch Cunnane (1978), S.211 oder Gumbel (1958), Kapitel 1.2.7.

Die Frage, welche Güte eine Schätzung eine hohen Quantils mit diesem Ansatz hat, beispielsweise des $1-\alpha$-Quantils $\mathrm{VaR}_\alpha(X)$ (Value at Risk) mit typischerweise kleinem $\alpha$, lässt sich ebenfalls leicht beantworten. Aus der Regressionsgeraden ergibt sich nämlich unmittelbar

$$\widehat{\mathrm{VaR}_\alpha(X)} = \hat{\mu}+\hat{\sigma}\cdot Q_Z(1-\alpha) \qquad (10)$$

mit

$$E\left(\widehat{\mathrm{VaR}_\alpha(X)}\right) = \mu+\sigma\cdot Q_Z(1-\alpha) = Q_X(1-\alpha), \qquad (11)$$

also ebenfalls eine erwartungstreue Schätzung.

Es gibt allerdings Probleme mit der Erwartungstreue, wenn die Daten vorher (z.B. logarithmisch) transformiert werden, um eine Lage-Skalen-Familie zu erhalten. Im letzterem Fall gilt für das Risiko $Y=\exp(X)$ nämlich $\mathrm{VaR}_\alpha(Y) = \mathrm{VaR}_\alpha\left(e^X\right) = e^{\mathrm{VaR}_\alpha(X)}$, die übliche Schätzung $\widehat{\mathrm{VaR}_\alpha(Y)} := e^{\widehat{\mathrm{VaR}_\alpha(X)}}$ besitzt aufgrund der Jensen-Ungleichung für konvexe Funktionen (vgl. etwa Czado und Schmidt (2011), Satz 1.5) aber einen Bias:

$$E\left(\widehat{\mathrm{VaR}_\alpha(Y)}\right) = E\left(e^{\widehat{\mathrm{VaR}_\alpha(X)}}\right) > e^{E\left(\widehat{\mathrm{VaR}_\alpha(X)}\right)} = e^{\mathrm{VaR}_\alpha(X)} = \mathrm{VaR}_\alpha(Y). \qquad (12)$$

Im Kontext von Solvency II ist das jedoch eher unschädlich, wenn die Schaden- bzw. Schaden-Kosten-Quoten als lognormalverteilt angenommen werden. Die logarithmierten Schaden- bzw. Schaden-Kosten-Quoten sind dann normalverteilt, so dass hier die Bestimmung bzw. numerische Berechnung der Erwartungswerte der Ordnungsstatistiken der Standard-Normalverteilung (mit Dichte $\varphi$ und Verteilungsfunktion $\Phi$) relevant ist. Formal gilt hier (vgl. David und Nagaraja (2003), Kapitel 3.1):

$$E\left(Z_{(k)}\right) = k\binom{n}{k}\int_{-\infty}^{\infty} x\,\Phi^{k-1}(x)\left(1-\Phi(x)\right)^{n-k}\varphi(x)\,dx\ \text{ für } k=1,\cdots,n. \qquad (13)$$

Dabei ist

$$\varphi(x) = \frac{1}{\sqrt{2\pi}}\exp\left(-\frac{x^2}{2}\right)\ \text{ und }\ \Phi(x) = \int_{-\infty}^{x}\varphi(u)\,du\ \text{ für } x\in\mathbb{R}. \qquad (14)$$

Sowohl $\Phi(x)$ als auch $E(Z_{(k)})$ sind nicht elementar berechenbar. Numerische Auswertungen für $E(Z_{(k)})$ wurden von Harter (1961) publiziert, für $n=1,\cdots,100$ und $k=1,\cdots,n$ vollumfänglich und für größere $n$ bis 400 in Auszügen. In der Literatur findet man dazu zahlreiche numerische Approximationen, z.B. (vgl. Cunnane (1978), S.211):

| Hazen (1914) | $E(Z_{(k)}) \approx \Phi^{-1}\left(\dfrac{k-0,5}{n+1}\right)$ |
|---|---|
| Weibull (1939) | $E(Z_{(k)}) \approx \Phi^{-1}\left(\dfrac{k}{n+1}\right)$ |
| Beard (1943) | $E(Z_{(k)}) \approx \Phi^{-1}\left(\dfrac{k-0,31}{n+0,38}\right)$ |
| Benard and Bos-Levenbach (1953) | $E(Z_{(k)}) \approx \Phi^{-1}\left(\dfrac{k-0,30}{n+0,20}\right)$ |
| Blom (1958) | $E(Z_{(k)}) \approx \Phi^{-1}\left(\dfrac{k-0,375}{n+0,25}\right)$ |
| Tukey (1962) | $E(Z_{(k)}) \approx \Phi^{-1}\left(\dfrac{k-0,333}{n+0,333}\right)$ |
| Gringorten (1963) | $E(Z_{(k)}) \approx \Phi^{-1}\left(\dfrac{k-0,44}{n+0,12}\right)$ |

Tab. 1: Approximationen für Erwartungswerte von Ordnungsstatistiken

Aus einer selbst durchgeführten größeren Monte-Carlo-Studie ergibt sich als sehr gute Approximation für die Erwartungswerte der Ordnungsstatistiken einer Standard-Normalverteilung:

$$E(Z_{(k)}) \approx \Phi^{-1}(\hat{u}_k) \text{ mit } \hat{u}_k = \frac{k-\hat{a}_n}{n+\hat{b}_n}, \qquad (15)$$

wobei

$$\hat{a}_n = 0,27950585 + \frac{0,04684273}{0,34986981 + n^{-0,79499457}},$$
$$\hat{b}_n = 0,44480354 - \frac{0,09890767}{0,36353365 + n^{-0,78493983}}, \ k=1,\ldots,n, \ n \leq 100. \qquad (16)$$

Der Fehler zwischen den exakten Werten für $a_n$ und $b_n$ (berechnet aus den Werten von Harter (1961)) und den approximativen Werten $\hat{a}_n$ und $\hat{b}_n$ gemäß (16) beträgt für $3 \leq n \leq 100$ jeweils betragsmäßig maximal 0,00007. Die folgenden Graphiken zeigen die exakten Werte von $a_n$ bzw. $b_n$ (Punkte) gegenüber den approximativen Werten $\hat{a}_n$ bzw. $\hat{b}_n$ (gestrichelte Linie).

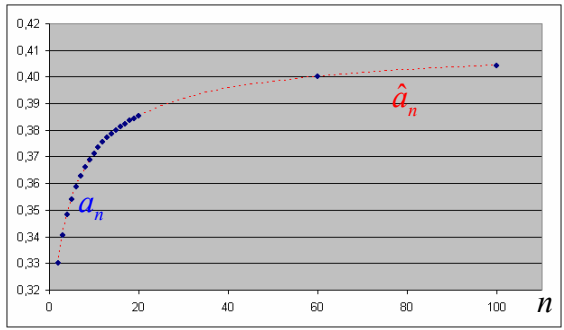 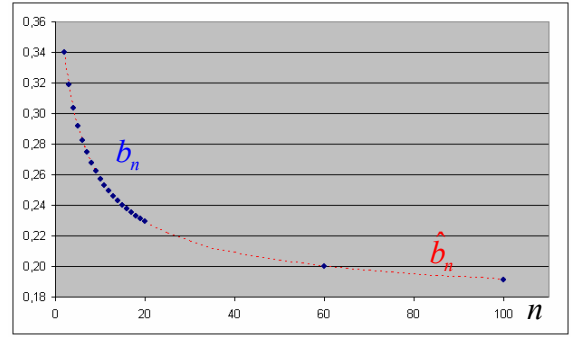

Abb.1: $a_n$ vs. $\hat{a}_n$  Abb. 2: $b_n$ vs. $\hat{b}_n$

Durch Vergleich mit den Werten aus Tab. 1 ergibt sich eine sehr gute Übereinstimmung zu Blom (1958) für $n=12$ und zu Tukey (1962) für $n=2$. Die übrigen Approximationsformeln weisen demgegenüber größere Abweichungen auf.

Für $n \leq 20$ kann auch die relativ gute Approximation

$$\hat{a}_n = 0{,}3177\, n^{0{,}0661}, \quad \hat{b}_n = \frac{0{,}3856}{n^{0{,}1754}}, \quad k = 1,\ldots, n \qquad (17)$$

verwendet werden mit einem betragsmäßig maximalen Fehler von 0,003.

## 3. Quantil-Quantil-Plots und Lage-Skalen-Familien: Tests

Es gibt in der Literatur eine Reihe von Vorschlägen zum Testen der Hypothese

$H_0$: die Verteilung des Risikos $X$ entstammt der Lage-Skalen-Familie zu $Z$

(sog. einfacher Signifikanztest). Eine gute Übersicht über dieses Thema geben Lockhart und Stephens (1998). Interessant sind hier Tests auf der Basis des empirischen Korrelationskoeffizienten $\rho_n$ aus dem Quantil-Quantil-Plot (Erwartungswerte der Ordnungsstatistiken gemäß (15) und (16) vs. den der Größe nach angeordneten Beobachtungswerten). Die Tatsache, dass die Verteilung des empirischen Korrelationskoeffizienten unter der Nullhypothese von den Lage- und Skalenparametern unabhängig ist, folgt aus der Invarianz der Korrelation zweier Risiken unter positiv-homogener (linearer) Transformation, vgl. Fahrmeir et al. (2016), Abschnitt 3.4.4. Ähnliche Testverfahren wurden u.a. von Shapiro und Wilk entwickelt, vgl. D'Agostino und Stephens (1986), Kapitel 5 oder Huber-Carol et al. (2002), Kapitel 7. Einen Vergleich über zahlreiche Anpassungstests auf Normalverteilung geben Seier (2002) und Yazici und Yolacan (2007).

In dieser Arbeit schlagen wir eine geringfügige Modifikation der in Lockhart und Stephens (1998) betrachteten Teststatistiken vor, nämlich $T_n := -\ln(1 - \rho_n)$. Dies hat den Vorteil, dass die Verteilung von $T_n$ unter der Nullhypothese einer Normalverteilung für Werte von $n \geq 10$ recht gut selbst durch eine Normalverteilung approximiert werden kann, so dass sich die $p$-Werte für den Anpassungstest einfach, z.B. mit EXCEL, berechnen lassen. (Zu den Grundlagen von *p*-Werten vgl. etwa Fahrmeir et al. (2016), Abschnitt 10.2.3.). Dieses Ergebnis deckt sich mit einer ähnlichen allgemeinen Aussage über die asymptotische Normalität des empirischen Korrelationskoeffizienten zwischen zwei Zufallsvariablen, vgl. etwa van der Vaart (1998), Example 3.6.

Das folgende Beispiel zeigt die Wirkungsweise bei einer Fehlerwahrscheinlichkeit 1. Art von 5%.

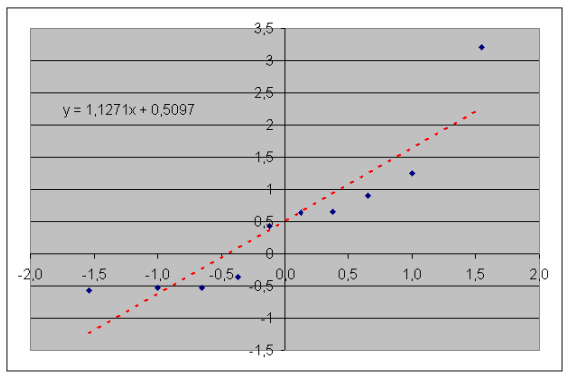 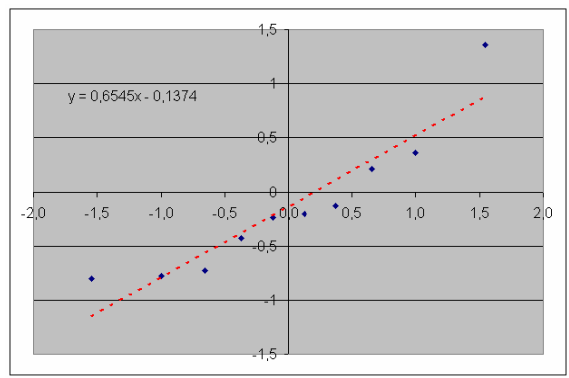

Abb. 3: $T_{10} = 2{,}4467;\ p = 4{,}22\%$      Abb. 4: $T_{10} = 2{,}7261;\ p = 10{,}26\%$

Nullhypothese wird abgelehnt      Nullhypothese wird angenommen

Die folgenden Graphiken zeigen die Histogramme aus jeweils 1 Mio. Simulationen der Verteilung der Teststatistik $T_n$ für $n = 11, \cdots, 20$ mit einer Anpassung an die Normalverteilung.

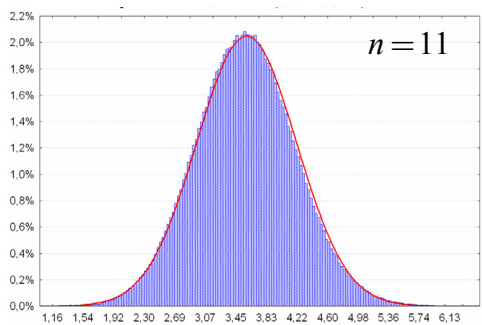 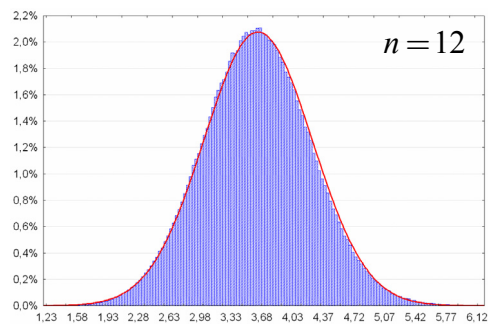

Abb. 5: $\mu = 3{,}5727;\ \sigma = 0{,}6201$      Abb. 6: $\mu = 3{,}6219;\ \sigma = 0{,}6103$

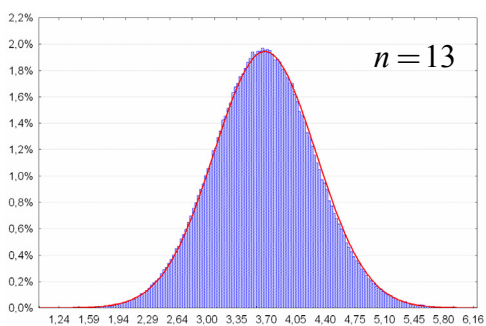 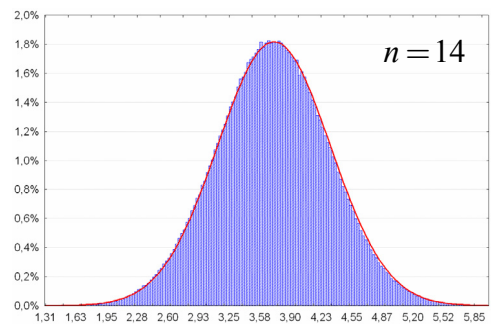

Abb. 7: $\mu = 3{,}6696;\ \sigma = 0{,}6005$      Abb. 8: $\mu = 3{,}7152;\ \sigma = 0{,}5935$

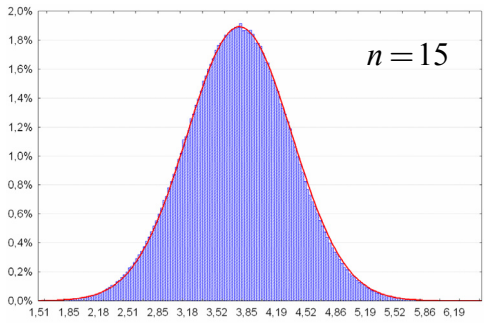

Abb. 9:  $\mu = 3{,}7584;\ \sigma = 0{,}5873$

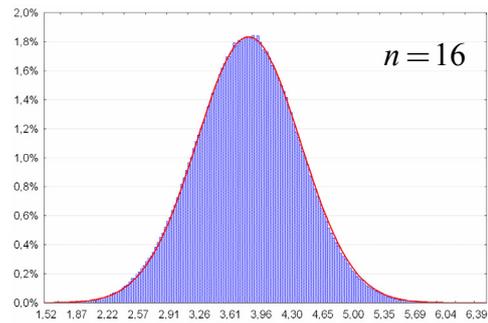

Abb. 10:  $\mu = 3{,}7998;\ \sigma = 0{,}5813$

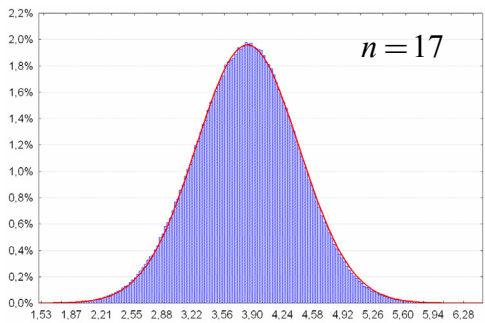

Abb. 11:  $\mu = 3{,}8385;\ \sigma = 0{,}5767$

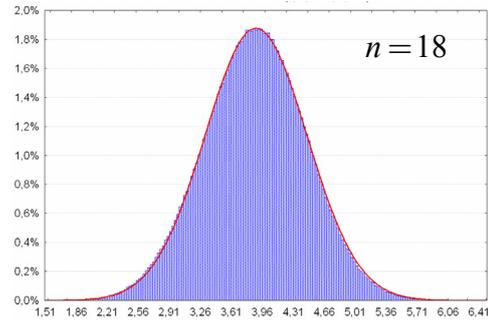

Abb. 12:  $\mu = 3{,}8773;\ \sigma = 0{,}5730$

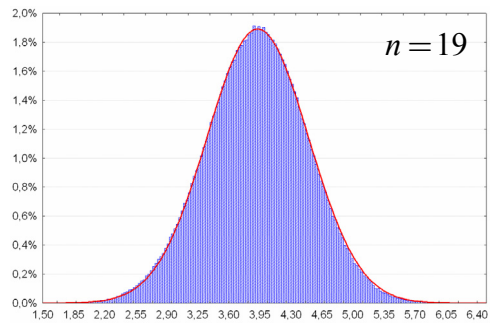

Abb. 13:  $\mu = 3{,}9119;\ \sigma = 0{,}5679$

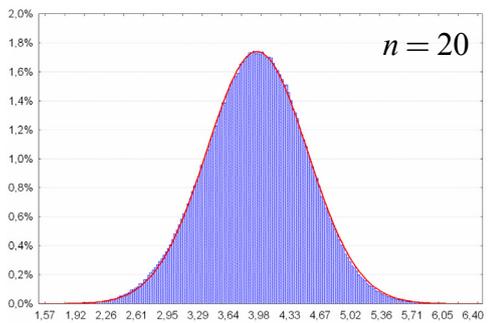

Abb. 14:  $\mu = 3{,}9475;\ \sigma = 0{,}5645$

Die Parameter $\mu_n$ und $\sigma_n$ der angepassten Normalverteilung lassen sich im Bereich $n = 10, \cdots, 50$ recht gut durch folgende Interpolation approximieren:

$$\hat{\mu}_n = \frac{5{,}87383\, n + 101{,}011}{n + 35{,}3404} \quad \text{und} \quad \hat{\sigma}_n = \frac{0{,}477812\, n + 3{,}25495}{n + 2{,}72721} \tag{18}$$

| $n$ | $\mu_n$ | $\hat{\mu}_n$ | $\sigma_n$ | $\hat{\sigma}_n$ |
|---|---|---|---|---|
| 10 | 3,5221 | 3,5233 | 0,6323 | 0,6312 |
| 11 | 3,5727 | 3,5741 | 0,6201 | 0,6200 |
| 12 | 3,6219 | 3,6226 | 0,6103 | 0,6103 |
| 13 | 3,6696 | 3,6692 | 0,6005 | 0,6019 |
| 14 | 3,7152 | 3,7139 | 0,5935 | 0,5945 |
| 15 | 3,7584 | 3,7568 | 0,5873 | 0,5879 |
| 16 | 3,7998 | 3,7980 | 0,5813 | 0,5820 |
| 17 | 3,8385 | 3,8377 | 0,5767 | 0,5768 |
| 18 | 3,8773 | 3,8759 | 0,5730 | 0,5720 |
| 19 | 3,9119 | 3,9126 | 0,5679 | 0,5676 |
| 20 | 3,9475 | 3,9481 | 0,5645 | 0,5637 |
| 26 | 4,1328 | 4,1364 | 0,5473 | 0,5458 |
| 50 | 4,6259 | 4,6250 | 0,5138 | 0,5148 |

Tab. 4: Parameterapproximation für Testverteilung

## 4. Gütebewertung

Die folgenden Graphiken zeigen im Vergleich Histogramme der Verteilung der Teststatistik für $n = 20$ bei den Hypothesen

$H_0$: die Verteilung des Risikos $X$ entstammt einer Normalverteilung

gegen

$H_1$: die Verteilung des Risikos $X$ entstammt einer Gumbelverteilung

bzw.

$H_2$: die Verteilung des Risikos $X$ entstammt einer logistischen Verteilung.

Der Testumfang betrug jeweils 1 Mio. Simulationen.

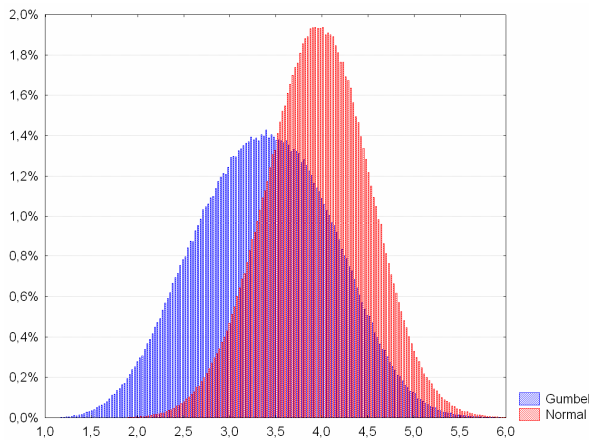 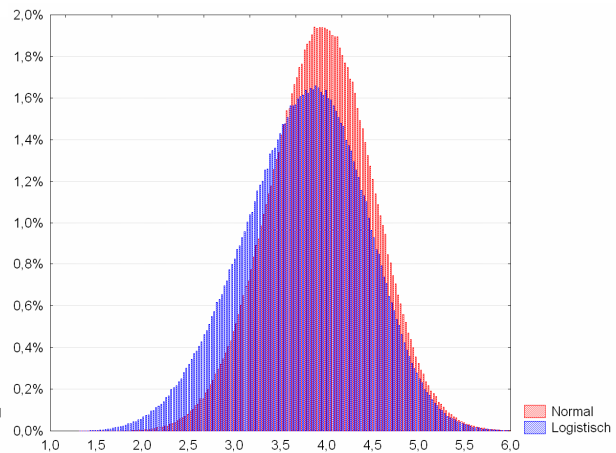

Abb. 15: Histogramme der Verteilung der Teststatistik unter $H_0$ (rot) und $H_1$ (blau)

Abb. 16: Histogramme der Verteilung der Teststatistik unter $H_0$ (rot) und $H_2$ (blau)

Erwartungsgemäß ist die Trennschärfe zwischen Normal- und Gumbelverteilung höher als zwischen Normal- und Logistischer Verteilung. Aus den Simulationen lassen sich die Fehlerwahrscheinlichkeiten zweiter ($\beta$) Art bei gegebener Fehlerwahrscheinlichkeit erster Art ($\alpha$) (approximativ) bestimmen:

| $\alpha$ | 1% | 5% | 10% |
|---|---|---|---|
| kritischer Wert | 2,6180 | 3,0045 | 3,2159 |
| $\beta$ Gumbel | 84,70% | 69,31% | 59,11% |
| $\beta$ Logistisch | 94,92% | 86,73% | 78,84% |

Tab. 5. Fehlerwahrscheinlichkeiten 2. Art für den Korrelationstest

Die Ergebnisse gelten nach Logarithmieren der Daten analog auch für das Testen einer Lognormalverteilung gegen eine Fréchet- bzw. Loglogistische Verteilung.

Es sollte hier noch angemerkt werden, dass der oben vorgestellte Korrelationstest eine im Allgemeinen bessere Trennschärfe besitzt als der Kolmogorov-Smirnov- bzw. Lilliefors-Test (vgl. Dallal und Wilkinson (1986)), wie auch schon von Durbin (1961) festgestellt wurde. Dies liegt daran, dass der Korrelationstest – bei gleichem Abstand der Verteilungen – empfindlicher auf Abweichungen in der Form der Verteilungsfunktionen reagiert. Die folgende Tabelle zeigt zum Vergleich exemplarisch die Fehlerwahrscheinlichkeiten zweiter Art für den Lilliefors-Test mit $n = 20$ bei den Alternativen Gumbel- und Logistische Verteilung auf. Die Daten stammen aus einer selbst durchgeführten Simulationsstudie mit einem jeweiligen Umfang von 1.000.000.

| $\alpha$ | 1% | 5% | 10% |
|---|---|---|---|
| kritischer Wert | 0,2230 | 0,1918 | 0,1762 |
| $\beta$ Gumbel | 92,15% | 79,60% | 69,84% |
| $\beta$ Logistisch | 97,66% | 91,40% | 84,85% |

Tab. 6. Fehlerwahrscheinlichkeiten 2. Art für den Lilliefors-Test

| $\alpha$ | 1% | 5% | 10% |
|---|---|---|---|
| kritischer Wert | 0,8672 | 0,9042 | 0,9199 |
| $\beta$ Gumbel | 84,09% | 68,76% | 58,50% |
| $\beta$ Logistisch | 95,94% | 88,38% | 81,90% |

Tab. 7. Fehlerwahrscheinlichkeiten 2. Art für den Shapiro-Wilk-Test

Für die Gumbel-Alternative sind die Fehlerwahrscheinlichkeiten zweiter Art hier nur minimal kleiner als bei dem Korrelationstest, für die Logistische Alternative etwas höher, allerdings deutlich kleiner als beim Lilliefors-Test.

## 5. Fallstudie

In diesem Abschnitt werden die im vorigen Teil beschriebenen Verfahren anhand von Informationen aus der Versicherungsbranche veranschaulicht. Konkret geht es um brutto-Schaden-Kosten-Quoten der Sparten Sach gesamt, Sach privat, verbundene Gebäude-versicherung (VGV), verbundene Hausratversicherung (VHV), Unfall, Rechtsschutz, Gewerbe und allgemeine Haftpflicht. Die Daten wurden dem vom GDV herausgegebenen Statistischen Taschenbuch der Versicherungsbranche (2018) ab dem Jahr 2000 entnommen (Haftpflicht ab 2004). Es soll geprüft werden, ob die Schaden-Kosten-Quoten lognormal-, also die logarithmierten Quoten normalverteilt sind. Der jeweilige Quantil-Quantil-Plot wird deshalb mit den logarithmierten Quoten erstellt. Die Nullhypothese lautet hier: die (logarithmierten) Quoten sind normalverteilt.

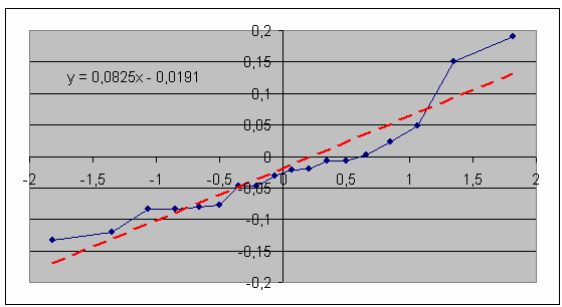 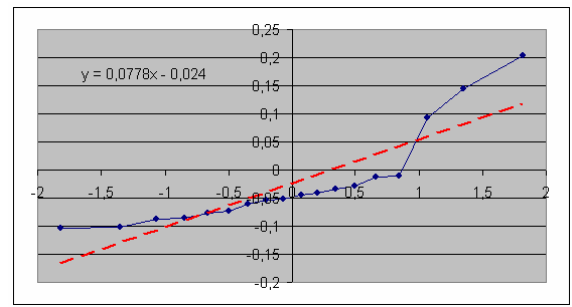

Abb. 17: Q-Q-Plot für Sach gesamt        Abb. 18: Q-Q-Plot für Sach privat

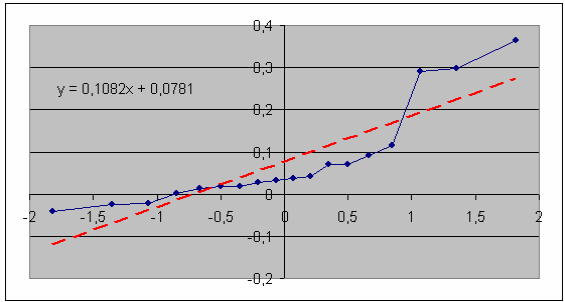
Abb. 19: Q-Q-Plot für VGV

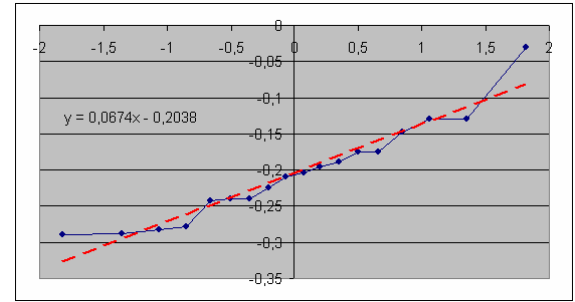
Abb. 20: Q-Q-Plot für VHV

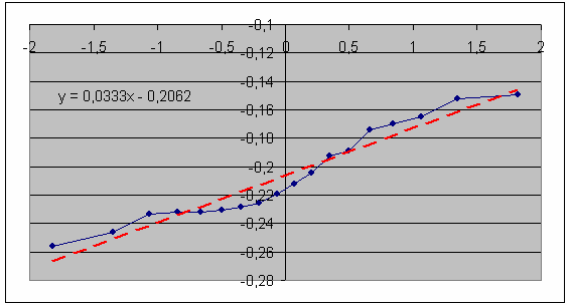
Abb. 21: Q-Q-Plot für Unfall

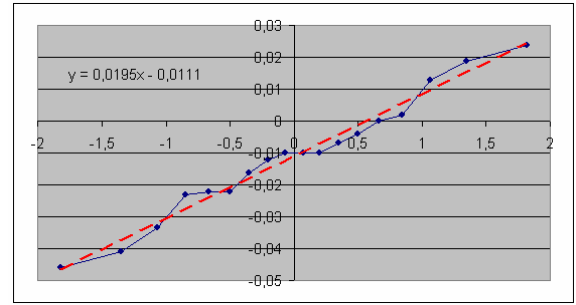
Abb. 22: Q-Q-Plot für Rechtsschutz

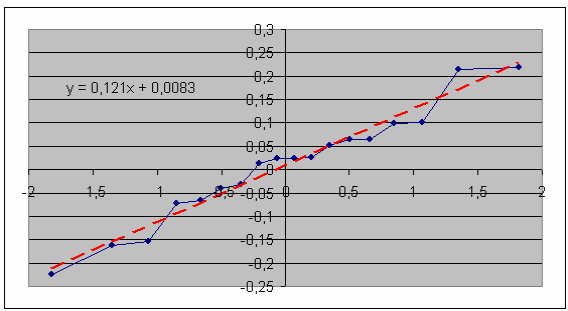
Abb. 23: Q-Q-Plot für Gewerbe

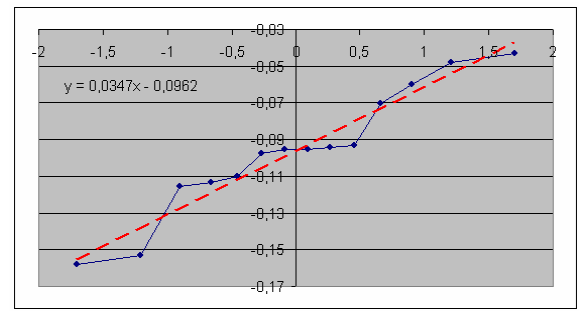
Abb. 24: Q-Q-Plot für allgemeine Haftpflicht

Die nachfolgende Tabelle enthält die Testgrößen $T_n$ und $p$-Werte für den Korrelationstest sowie zum Vergleich die Testgrößen $d_n$ und $W_n$ sowie $p$-Werte nach dem Lilliefors- bzw. Shapiro-Wilk-Test. Die $p$-Werte für die Teststatistiken $d_n$ und $W_n$ wurden durch eine eigene Monte Carlo Studie im Umfang von 1.000.000 ermittelt, vgl. auch Dallal und Wilkinson (1986).

|  | Korrelationstest | | Lilliefors-Test | | Shapiro-Wilk-Test | |
|---|---|---|---|---|---|---|
| Sparte | $T_n$ | $p$-Wert | $d_n$ | $p$-Wert | $W_n$ | $p$-Wert |
| Sach gesamt | 2,8831 | 4,32% | 0,1743 | 15,17% | 0,8955 | 9,67% |
| Sach privat | 2,1064 | 0,09% | 0,2732 | 0,11% | 0,7758 | 0,20% |
| VGV | 2,1378 | 0,11% | 0,2481 | 0,47% | 0,7785 | 0,21% |
| VHV | 3,3515 | 17,95% | 0,1083 | 83,00% | 0,9329 | 33,18% |
| Unfall | 3,5261 | 26,76% | 0,1670 | 19,73% | 0,9280 | 28,50% |
| Rechtsschutz | 4,6539 | 91,30% | 0,0936 | 94,59% | 0,9763 | 91,62% |
| Gewerbe | 4,1377 | 67,08% | 0,1268 | 61,19% | 0,9665 | 79,25% |
| Allgemeine Haftpflicht | 3,9443 | 64,96% | 0,1795 | 24,90% | 0,9395 | 40,59% |

Tab. 8: Testgrößen und $p$-Werte

Es zeigt sich, dass sich die $p$-Werte aller Verfahren deutlich unterscheiden, mit der größten Diskrepanz bei den Sparte Sach gesamt, VHV und Allgemeine Haftpflicht. Bei den Sparten Sach privat und VGV lehnen alle Tests bei einer Fehlerwahrscheinlichkeit 1. Art von 1% die Nullhypothese ab. Bei der Sparte Sach gesamt würde die Nullhypothese bei einer Fehlerwahrscheinlichkeit 1. Art von 5% mit dem Korrelationstest knapp abgelehnt, mit den anderen beiden Tests dagegen nicht. Bei den Sparten VHV, Unfall, Rechtsschutz, Gewerbe und Allgemeine Haftpflicht wird die Nullhypothese bei einer Fehlerwahrscheinlichkeit 1. Art von 10% unter allen Tests nicht verworfen.

Als geschätzte Parameter für die Normal- bzw. Lognormalverteilung erhält man jeweils aus dem Achsenabschnitt den geschätzten Erwartungswert $\hat{\mu}$ und aus der Steigung die geschätzte Streuung $\hat{\sigma}$.

## 6. Fazit

Ein Korrelationstest auf der Basis von Quantil-Quantil-Plots ist einfach durchzuführen und hat gegenüber dem Lilliefors- und anderen Anpassungstests den Vorteil, explizit näherungsweise gute $p$-Werte aller Größenordnungen für beliebige Stichprobenumfänge zu erhalten, mit einer guten Anpassung der Verteilung der Testgröße an eine Normalverteilung. Empirische Studien zeigen dabei eine vergleichbare Güte zu ähnlichen Testverfahren.

Quantil-Quantil-Plots bieten darüber hinaus den großen Vorteil einer graphischen Veranschaulichung der Testergebnisse, was insbesondere für mathematisch weniger geschulte Mitarbeiter von Versicherungsunternehmen interessant sein dürfte.

# Literatur